\documentclass[11pt]{article}

\usepackage{amssymb,amsmath,amsthm}
\usepackage{authblk}
\usepackage{multirow}
\usepackage{blindtext}
\usepackage{amsfonts}
\usepackage{amsmath,amssymb,pxfonts,xcolor}
\usepackage{slashed}
\usepackage{graphicx}
\usepackage{comment}
\usepackage{ulem}
\usepackage{color}
\usepackage{pgfplots}
\pgfplotsset{compat=1.9}

\newtheorem{theorem}{Theorem}

\title{A lower bound for the volatility swap in the lognormal SABR model}
\author{Elisa Al\`os\thanks{Universitat Pompeu Fabra and Barcelona Graduate School of Economics. Suported by grant  PID2020-118339GB-I00  ( Ministerio de Ciencia e Innovaci\'on)} \quad Frido Rolloos\thanks{Independent researcher.} \quad Kenichiro Shiraya\thanks{Graduate School of Economics, The University of Tokyo. Supported by CARF.}}

\begin{document}
\maketitle

\abstract{In the short time to maturity limit, it is proved that for the conditionally lognormal SABR model, the zero vanna implied volatility is a lower bound for the volatility swap strike. The result is valid for all values of the correlation parameter and is a sharper lower bound than the at-the-money implied volatility for correlation less than or equal to zero.}

\section{Introduction}
In their seminal work Carr and Lee \cite{CL1} prove that for stochastic volatility models, with zero correlation between the spot price and its instantaneous volatility, general volatility derivatives can be priced in a model-free manner. This is a surprising and important result as prior to Carr and Lee's result it was believed that volatility derivatives other than the log-contract are highly model-dependent. Indeed, precisely because of its model independence one of the first volatility derivatives to be traded in the market was the variance swap, which is equivalent to the log-contract.

When correlation is non-zero prices of general volatility derivatives do become model-dependent. Nevertheless, Carr and Lee proved that the model-dependence is of the order $O(\rho^2)$. Theoretically, their result paves the way for pricing and replicating all European type volatility derivatives with an error of $O(\rho^2)$. However, as shown by Friz and Gatheral \cite{FG} even in the zero correlation case, pricing volatility options using the Carr and Lee immunisation method leads to highly oscillatory integrals which need regularisation techniques to evaluate. Furthermore, even for the relatively simple but important volatility swap, the Carr-Lee's pricing method requires a continuum of vanilla options (strikes from $0$ to $\infty$) . 

A more recent and straightforward approximation for the volatility swap is given by Rolloos and Arslan \cite{RA}. In the aforementioned paper, it is shown that the so-called zero vanna implied volatility (ZVIV) approximates the volatility swap price. In contrast to the Carr-Lee method, the ZVIV approximation does not require a continuum of options, but only one option. It therefore simplifies the approximate pricing of volatility swaps and is not prone to extrapolation error.

The error between the exact volatility swap price and the ZVIV in the short time to maturity limit was made precise by Al\`os et al. \cite{ARS} for a class of models that include in particular the lognormal SABR, by making use of techniques from the Malliavin calculus. In their paper Al\`os et al. quantify the error for general stochastic volatility models driven by fractional noise. Recent studies \cite{CR,GJR} have argued for the relevance of fractional noise in explaining the shape of the implied volatility skew.

In Al\`os et al. numerical simulations also indicate that, in addition to approximating volatility swaps, the ZVIV is a lower bound for the volatility swap. In this paper it is proved that for the lognormal SABR model (henceforth SABR model), this is indeed the case for all values of correlation in the short time to maturity limit. To the best of our knowledge this has not been proved yet in the literature on derivatives pricing. A proof for general SV models is a subject of future research.

The structure of the papaer is as follows. In Section 2, the working assumptions are stated and relevant notation introduced. Section 3 treats the zero correlation case. It is proved that both the at-the-money implied volatility (ATMI) and the ZVIV are lower bounds for the volatility swap in the short maturity limit in the SABR model. By making use of the result that the ZVIV is a better approximation than the ATMI, it follows that the ZVIV is a sharper lower bound. The nonzero correlation case is discussed in Section 4, wherein it is proved that in the SABR model the ZVIV is a lower bound for the volatility swap price for any value of correlation. Furthermore, it is shown that for non-positive correlation it is a sharper lower bound than the ATMI in the short time to maturity limit. To illustrate the different behaviours of the ATMI and the ZVIV, numerical results in Section 5 are presented for various values of correlation, volatility of volatility and time to maturity. Section 6 concludes.

\section{Assumptions and notation}

We consider the log normal stochastic volatility model for the log-price of a stock
under a risk-neutral probability measure $P$:
\begin{align}
\label{themodel}
X_{t} &=X_0-\frac{1}{2}\int_{0}^{t}{\sigma _{s}^{2}}ds+\int_{0}^{t}\sigma
_{s}\left( \rho dW_{s}+\sqrt{1-\rho ^{2}}dB_{s}\right), \\
\label{SABR}
\sigma_t &= \sigma_0 + \int_0^t \alpha \sigma_s \, dW_s,
\end{align}
where $X_0$ is the current log-price, and $W$ and $B$ are independent standard Brownian motions defined on a complete probability
space  $(\Omega ,\mathcal{G},P)$.  The initial volatility $\sigma_0$ and the volatility of volatility $\alpha$ are  positive constants. We denote by $\mathcal{F}^{W}$ and $\mathcal{F}^{B}$
the filtrations generated by $W$ and $B$ and $\mathcal{F}:=%
\mathcal{F}^{W}\vee \mathcal{F}^{B}.$ We assume the risk-free interest rate $r$ is zero for simplicity. The same arguments of the results in this paper hold for $r\neq 0$.

Under this model, the value of the European call option with strike price $e^k$ is
\[
V_{t}=E_{t}[(e^{X_{T}}-e^k)_{+}], 
\]%
where $E_{t}$ denotes the $\mathcal{F}_{t}-$conditional expectation with respect
to $P$ (that is, $E_{t}[Z]=E[Z|\mathcal{F}_{t}]$). 
Also, we use the following notations:

\begin{itemize}
\item $v_t=\sqrt{\frac{Y_t}{T-t}}$, where $Y_t=\int_{t}^{T}\sigma _{u}^{2}du$.

$v$ is the future average volatility and is not an adapted process. 
$E_{t}\left[
v_{t}\right] $ is the fair strike of a volatility swap with the time to maturity time $T-t$.

\item $BS(t,T,x,k,\sigma )$ is the European call option price under the Black-Scholes model with the constant volatility $\sigma $,
the stock price $e^x$, the time to maturity $T-t,$ and the strike $e^k $. 
In this setting,
\[
BS(t,T,x,k,\sigma )=e^{x}N(d_1(k,\sigma ))-e^{k}N(d_2(k,\sigma )), 
\]%
where $N$ is the cumulative distribution function of the standard normal distribution, and
\[
d_1\left( k,\sigma \right) :=\frac{x-k}{\sigma \sqrt{T-t}}+
\frac{\sigma }{2}\sqrt{T-t},  \hspace{0.4cm} d_2\left( k,\sigma \right) :=\frac{x-k}{\sigma \sqrt{T-t}}-
\frac{\sigma }{2}\sqrt{T-t}.
\]%
We use the notation $BS(k,\sigma):=BS(t,T,X_t,k,\sigma )$ for simplicity.
\item  The inverse function $BS^{-1}(t,T,x,k, \cdot)$ of the Black-Scholes European call option pricing formula with respect to the volatility is defined as
\[
BS(t,T,x,k, BS^{-1}( t,T,x,k,\lambda) )=\lambda,
\]
for all $\lambda>0$.
We use the notation $BS^{-1}(k,\lambda)\ :=BS^{-1}(t,T,X_{t},k,\lambda)$ for simplicity.

\item For any fixed $t,T,X_{t},k,$ we define the implied volatility function $I(t,T,X_{t},k) $ as
\[
BS( t,T,X_{t},k,I( t,T,X_{t},k) ) =V_{t}.
\]
Then, $I(t,T,X_t,k)=BS^{-1}( t,T,X_t,k,V_t)$.

\item $k_t^*$ denotes the ATM strike at time $t$, and $I(t,T,X_t,k_t^*)$ is the corresponding ATMI

\item $\hat{k_t}$ is the {\it zero vanna strike} at time $t$ and satisfies $$d_2(\hat{k}_t,I(t,T,X_t,\hat{k}_t))=0.$$ 
$I(t,T,X_t,\hat{k}_t)$ is called {\it zero vanna implied volatility}. The Black-Scholes vanna is proportional to $d_2$, and that of plain vanilla options is zero at the zero vanna strike.
\end{itemize}

$\mathbb{D}_{W}^{1,2}$ denotes the domain of the
Malliavin derivative operator $D^{W}$ with respect to $W$.
For $n>1,$ the domains of the iterated derivatives $D^{n,W}$ are denoted by 
$\mathbb{D}_{W}^{n,2}$. Also, we define $\mathbb{L}_{W}^{n,2}=$\ $L^{2}(\left[ 0,T\right] ;\mathbb{D}_{W}^{n,2})$. Notice that the SABR variance $\sigma^2$ given by (\ref{SABR}) is a process in $\mathbb{L}_{W}^{2,2}$, and 
\begin{eqnarray}
D_r\sigma_u^2 &=& 2\alpha \sigma_u^2 1_{[0,u)}(r),\\
D_s D_r \sigma_u^2 &=& 2 D_s\sigma_u D_r\sigma_u + 2\sigma_u D_s D_r \sigma_u\nonumber\\
&=& 2 \alpha^2 \sigma_u^2 1_{[0,u)}(r \vee s) + 2 \alpha^2 \sigma_u^2 1_{[0,u)}(r \vee s)\nonumber\\
&=& 4 \alpha^2 \sigma_u^2 1_{[0,u)}(r \vee s).
\end{eqnarray}

\section{The uncorrelated case}

\subsection{ATM implied volatility}
\begin{theorem} Consider the SABR model defined as in (\ref{themodel}) and (\ref{SABR}). Then, for zero correlation, the ATMI is a lower bound for the volatility swap in the short time to maturity limit.
\begin{equation}
\lim_{T \to t} \, I( t,T,X_t,k^*_t) \leq \lim_{T \to t} \,  E_{t}[v_{t}].
\end{equation}
\begin{proof}
This follows immediately from Theorem 3.2 of Al\`os and Shiraya \cite{AS},
\begin{eqnarray}
\lim_{T\rightarrow t}\frac{I\left( t,T,X_t,k^*_t\right) -E_{t}\left[v_{t}\right] }{(T-t)^{2}}
= -\frac{1}{32\sigma _{t}}\lim_{T\rightarrow t}
\frac{1}{(T-t)^{3}}E_{t}\left[\int_{t}^{T}\left(E_{r}\left[\int_{r}^{T}D_{r}^{W}\sigma_{s}^{2}ds\right]\right) ^{2}dr\right] \le 0.
\end{eqnarray}
\end{proof}
\end{theorem}

\subsection{Zero-vanna implied volatility}

\begin{theorem}Consider the SABR model defined as in (\ref{themodel}) and (\ref{SABR}). Then, for zero correlation, the ZVIV is a lower bound for the volatility swap in the short time to maturity limit.
\begin{equation}
\lim_{T \to t} \, I( t,T,X_t,\hat{k}_t) \leq \lim_{T \to t} \,  E_{t}[v_{t}].
\end{equation}
\begin{proof}
In the uncorrelated case, since (B.3) of Al\`os, Rolloos and Shiraya,
\begin{eqnarray}
\label{primerpas}
I\left( t,T,X_t,\hat{k}_t\right) &=& E_{t}\left[ v_{t}\right] -  E_t\Bigg[\int_{t}^{T}\left( BS^{-1}\left( \hat{k}_t, \Lambda_r\right)\right) ^{\prime \prime\prime} (D^-A)_r U_r dr\Bigg]\nonumber\\
&&- \frac12 E_t\Bigg[
\int_{t}^{T}\left( BS^{-1}\left( \hat{k}_t, \Lambda_r\right)\right) ^{(iv)} A_r U_r^2 dr \Bigg]
\label{I-E[v]}\nonumber\\
&=& 
E_{t}\left[ v_{t}\right] -\left( BS^{-1}\left( \hat{k}_t, \Lambda_t\right)\right) ^{\prime \prime\prime} E_t\Bigg[\int_{t}^{T}(D^-A)_r U_r dr\Bigg]\nonumber\\
&&+ \frac12 \left( BS^{-1}\left( \hat{k}_t, \Lambda_t\right)\right) ^{(iv)} E_t\Bigg[
\int_{t}^{T} A_r U_r^2 dr \Bigg]\nonumber\\
&&+o((T-t)^{4H+1})\nonumber\\
&=& E_{t}\left[ v_{t}\right] + T_1 + T_2 +o((T-t)^{3}),
\end{eqnarray}
where 
\begin{eqnarray}
\Lambda _{r}&:=&E_{r}\left[ BS\left( t,T,X_t,\hat{k}_t,v_{t}\right) \right],\\
A_r&:=&\frac12 \int_{r}^{T} U_{s}^{2}ds,\\
(D^-A)_r&:=&\frac{1}{2}\int^T_r D_r^W U^2_s ds,\\
U_{r}&:=&E_{r}\left[D_r^W\left(BS(t,T,X_t,\hat{k}_t,v_t)\right)\right]\nonumber\\
&=&E_r\left[\frac{\partial BS}{\partial \sigma}(t,T,X_t,\hat{k}_t,v_t)\frac{1}{2v_t(T-t)}\int_r^T D_r^W\sigma_s^2 ds\right].
\end{eqnarray}
Here,
\begin{eqnarray}
\label{terceraderivada}
\left( BS^{-1}\left( \hat{k}_t, \Lambda_t\right)\right) ^{\prime \prime\prime}
&=&(2\pi)^{\frac32}\exp\left(-3X_t+\frac{3}{2}(\Theta_t(\hat{k}_t))^2(T-t)\right)(T-t)^{-\frac12}+o\left((T-t)^{-\frac{1}{2}}\right),\\
\label{derivada4}
\left( BS^{-1}\left( \hat{k}_t, \Lambda_t\right)\right) ^{(iv)}&=&-\frac{3(2\pi)^{2}}{\Theta_t(\hat{k}_t)}\exp\left(-4X_t+2(\Theta_t(\hat{k}_t))^2(T-t)\right)(T-t)^{-1} + o\left((T-t)^{-1}\right),
\end{eqnarray}
where $\Theta_{r}(k):=BS^{-1}(k,\Lambda_r)$.
Thus, $(BS^{-1})'''>0$ and $(BS^{-1})^{(iv)}<0$ in the short time limit.

On the other hand, 
\begin{eqnarray}
\lefteqn{E_t\Bigg[\int_{t}^{T} A_r U_r^2 dr \Bigg]}\nonumber\\
&=&\frac{1}{2}E_t\Bigg[\int_{t}^{T} \left(\int_r^TU_s^2ds\right) U_r^2 dr \Bigg]\nonumber\\
&=&\frac{1}{4}E_t\Bigg[ \left(\int_t^TU_r^2dr\right)^2  \Bigg]\nonumber\\
&=&\frac{1}{4}E_t\Bigg[ \left(\int_t^T\left(    E_r\left[\frac{\partial BS}{\partial \sigma}(t,T,X_t,\hat{k}_t,v_t)\frac{1}{2v_t(T-t)}\int_r^TD_r^W\sigma_s^2 dr\right]
    \right)^2ds\right)^2  \Bigg]\nonumber\\
&>& 0.
\end{eqnarray}
Therefore, $T_2\le 0$.

Next,
\begin{eqnarray}
\label{U}
U_s&=&E_s\left[\frac{\partial BS}{\partial \sigma}(t,T,X_t,\hat{k}_t,v_t)\frac{1}{2v_t(T-t)}\int_s^T D_s^W \sigma_u^2 du\right]  \nonumber\\
&=&\frac12    E_s\left[G(t,T,X_t,\hat{k}_t,v_t)\int_s^T D_s^W \sigma_u^2 du\right],
\end{eqnarray}
where $G(t,T,x,k,\sigma ):=( \frac{\partial ^{2}}{\partial x^{2}}-\frac{\partial}{\partial x}) BS(t,T,x,k,\sigma )$,
and
\begin{eqnarray}
\label{DU}
D_r^WU_s&=& E_s\Bigg[ \frac12 
G(t,T,X_t,\hat{k}_t,v_t)\left(\frac{d_1(\hat{k}_t,v_t)d_2(\hat{k}_t,v_t)}{2v_t^2(T-t)} - \frac{1}{2v_t^2(T-t)}\right)
\left(\int_s^T D_s^W\sigma_u^2 du\right)\left(\int_r^TD_r^W\sigma_u^2 du\right)
\nonumber\\
&&+\frac12G(t,T,X_t,\hat{k}_t,v_t)\left(\int_s^TD_r^WD_s^W\sigma_u^2 du\right)\Bigg],
\end{eqnarray}
\begin{eqnarray}
\label{DA2}
(D^-A)_r
&=&
\frac{1}{4}
\int_r^T E_s\left[\frac{e^{\hat{k}_t}N'(d_2(\hat{k}_t,v_t))}{v_t\sqrt{T-t}}\int_s^TD_s^W\sigma_u^2 du    \right] \nonumber\\
&&\times E_s \Bigg[
\frac{e^{\hat{k}_t}N'(d_2(\hat{k}_t,v_t))}{v_t\sqrt{T-t}}
\left(
\frac{-1}{2 v_t^2(T-t)}
\left(\int_s^TD_s^W\sigma_u^2 du\right)\left(\int_r^TD_r^W\sigma_u^2 du\right)+\left(\int_{r\vee s}^TD_r^WD_s^W\sigma_u^2 du\right)\right)
\Bigg] ds.\nonumber\\
&&
\end{eqnarray}
Thus, we need to check
\begin{equation}
\frac{-1}{2 v_t^2(T-t)}\int_s^TD_s^W\sigma_u^2 du\int_r^TD_r^W\sigma_u^2 du+\int_s^TD_r^WD_s^W\sigma_u^2 du \ge 0.
\label{eq9}
\end{equation}
Since $\int_t^T \sigma_u^2 du \ge \int_r^T \sigma_u^2 du$, \eqref{eq9} is
\begin{eqnarray}
\lefteqn{\frac{-1}{2 \int_t^T \sigma_u^2 du}\int_s^T 2\alpha \sigma_u^2 du\int_r^T 2\alpha \sigma_u^2 du+\int_{r\vee s}^T 4\alpha^2 \sigma_u^2 du} \nonumber\\
&\ge& - \int_{r\vee s}^T 2\alpha^2 \sigma_u^2 du+\int_{r\vee s}^T 4\alpha^2 \sigma_u^2 du \nonumber\\
&\ge& 0.
\end{eqnarray}
Therefore, $T_1 \le 0$ and the zero-vanna iv is a lower bound for the volatility swap.
\end{proof}
\end{theorem}

\section{The correlated case}

\subsection{ATM implied volatility}

\begin{theorem}\label{correlatedATMI}
Consider the SABR model defined as in (\ref{themodel}) and (\ref{SABR}). Then, for $\rho < 0$, the ATMI is a lower bound for the volatility swap strike in the short time to maturity limit.
\begin{equation}
\lim_{T \to t} \, I(t,T,X_t,k^*_t) \leq \lim_{T \to t} \,  E_{t}[v_{t}].
\end{equation}
\begin{proof}
From Theorem 4.2 in \cite{AS},
\begin{eqnarray}
\label{lim1}
\lefteqn{\lim_{T\rightarrow t}\frac{I( t,T,X_t,{k}_t^*) -E_{t}[v_{t}]} {(T-t)}}\nonumber \\
&=& \lim_{T\rightarrow t}\frac{3\rho^{2}}{8\sigma _{t}^{3}(T-t)^{4}}
E_{t}\left[\left(\int_{t}^{T}\int_{s}^{T}D_{s}^{W}\sigma _{r}^{2}dr ds\right)^{2} \right]\nonumber\\
&&-\lim_{T\rightarrow t}\frac{\rho ^{2}}{{2\sigma _{t}^2(T-t)^{3}}}
E_{t}\left[ \int_{t}^{T} \int_{s}^{T}D_{s}^{W}\sigma_{r} \int_{r}^{T}D_{r}^{W}\sigma _{u}^{2}du dr ds\right]\nonumber\\
&&-\lim_{T\rightarrow t}\frac{\rho ^{2}}{2\sigma _{t}(T-t)^{3}}
E_{t}\left[\int_{t}^{T} \int_{s}^{T}\int_{r}^{T}D_{s}^{W}D_{r}^{W}\sigma _{u}^{2}du dr ds\right]\nonumber\\
&&+\lim_{T\rightarrow t}\frac{\rho}{4(T-t)^{2}}
E_{t}\left[\int_{t}^{T}\int_{s}^{T}D_{s}^{W}\sigma _{r}^{2}dr ds \right]\nonumber\\
&=& \lim_{T\rightarrow t}\frac{3\rho^{2}}{8\sigma _{t}^{3}(T-t)^{4}}T_1 - \lim_{T\rightarrow t}\frac{\rho ^{2}}{{2\sigma _{t}^2(T-t)^{3}}} T_2 - \lim_{T\rightarrow t}\frac{\rho ^{2}}{2\sigma _{t}(T-t)^{3}} T_3 +\lim_{T\rightarrow t}\frac{\rho}{4(T-t)^{2}} T_4.
\end{eqnarray}

In the log normal stochastic volatility model, since Schwartz's inequality,
\begin{eqnarray}
T_1 
&\le& 
(T-t)\int_{t}^{T}(T-s)\int_{s}^{T}E_{t}\left[\left(D_{s}^{W}\sigma _{r}^{2}\right)^{2}\right] dr ds \nonumber\\
&=&
(T-t)\int_{t}^{T}(T-s)\int_{s}^{T}4\alpha^2 E_{t}\left[ \sigma_0^4 e^{4\alpha W_r - 2\alpha^2 r}\right] dr ds \nonumber\\
&=&
(T-t)\int_{t}^{T}(T-s)\int_{s}^{T}4\alpha^2 \sigma_0^4 e^{6 \alpha^2 r} dr ds \nonumber\\
&=&
4\alpha^2 \sigma_0^4 (T-t)\int_{t}^{T}(T-s)\frac{e^{6 \alpha^2 T} - e^{6 \alpha^2 s}}{6\alpha^2} ds \nonumber\\
&=&
4\alpha^2 \sigma_0^4  \left(\frac{e^{6 \alpha^2 T}(T-t)^3}{12\alpha^2} + \frac{e^{6 \alpha^2 t}}{36\alpha^4}(T-t)^2 - (T-t)\int_t^T \frac{e^{6 \alpha^2 s}}{36\alpha^4}ds \right)\nonumber\\
&=&
4\alpha^2 \sigma_0^4  \left(\frac{e^{6 \alpha^2 T}(T-t)^3}{12\alpha^2} + \frac{e^{6 \alpha^2 t}}{36\alpha^4}(T-t)^2 - \frac{e^{6 \alpha^2 T} - e^{6 \alpha^2 t}}{216\alpha^6}(T-t) \right).\label{t1}
\end{eqnarray}
Here, we assume $t=0$ and apply the Taylor expansion,
\begin{eqnarray}
\eqref{t1}
&=& 
4\alpha^2 \sigma_0^4  \left(\frac{1+6\alpha^2T }{12\alpha^2}T^3 + \frac{T^2}{36\alpha^4} - \frac{6\alpha^2T + \frac{1}{2}36\alpha^4 T^2 + \frac{1}{6}216\alpha^6T^3}{216\alpha^6}T\right) + O(T^5)\nonumber\\
&=& 
4\alpha^2 \sigma_0^4  \left(\frac{1}{2}T^4 - \frac{1}{6}T^4\right) + O(T^5)\nonumber\\
&=& 
\frac{4 \alpha^2 \sigma_0^4}{3}T^4+O(T^5).
\end{eqnarray}

\begin{eqnarray}
T_2 
&=& \int_{t}^{T} \int_{s}^{T} \int_{r}^{T} E_{t}\left[ D_{s}^{W}\sigma_{r}  D_{r}^{W}\sigma _{u}^{2} \right]du dr ds\nonumber\\
&=& \int_{t}^{T} \int_{s}^{T} \int_{r}^{T} 2\alpha^2 \sigma_0^3  E_{t}\left[ e^{2\alpha(W_u - W_r) - 2\alpha^2(u-r)}\right]E_{t}\left[ e^{3\alpha W_r - \frac12 9\alpha^2r}\right]e^{\alpha^2u + 2\alpha^2r}du dr ds\nonumber\\
&=& 2\alpha^2 \sigma_0^3 \int_{t}^{T} \int_{s}^{T} \int_{r}^{T} e^{\alpha^2u + 2\alpha^2r}du dr ds\nonumber\\
&=& 2\alpha^2 \sigma_0^3 \int_{t}^{T} \int_{s}^{T} \frac{e^{\alpha^2T + 2\alpha^2r} - e^{3\alpha^2r}}{\alpha^2} dr ds\nonumber\\
&=& 2\alpha^2 \sigma_0^3 \int_{t}^{T} \frac{e^{3\alpha^2T} - e^{\alpha^2T + 2\alpha^2s}}{2\alpha^4} - \frac{e^{3\alpha^2T} - e^{3\alpha^2s}}{3\alpha^4} dr ds\nonumber\\
&=& 2\alpha^2 \sigma_0^3 \left( \frac{e^{3\alpha^2T}(T-t)}{2\alpha^4} - \frac{e^{3\alpha^2T} - e^{\alpha^2T + 2\alpha^2t}}{4\alpha^6} - \frac{e^{3\alpha^2T}(T-t)}{3\alpha^4} + \frac{e^{3\alpha^2T} - e^{3\alpha^2t}}{9\alpha^6} \right).\label{t2}
\end{eqnarray}
Here, we assume $t=0$ and apply the Taylor expansion,
\begin{eqnarray}
\eqref{t2}
&=& 
2\alpha^2 \sigma_0^3  \Bigg(\frac{(1+3\alpha^2T+\frac{1}{2}9\alpha^4T^2)T}{6\alpha^4} - \frac{(2\alpha^2T + \frac{1}{2}4\alpha^4T^2 + \frac{1}{6}8\alpha^6T^3)(1 + \alpha^2T + \frac{1}{2}\alpha^4T^2)}{4\alpha^6}\nonumber\\
&& + \frac{3\alpha^2 T+ \frac{1}{2}9\alpha^4T^2+\frac{1}{6}27\alpha^6T^3}{9\alpha^6}\Bigg) +O(T^4)\nonumber\\
&=&
\frac{\alpha^2 \sigma_0^3}{3}T^3+O(T^4).
\end{eqnarray}

\begin{eqnarray}
T_3 
&=& \int_{t}^{T} \int_{s}^{T}\int_{r}^{T} E_{t}\left[ 4 \alpha^2\sigma_0 e^{2\alpha W - \alpha^2 u}1_{[0,u)}(r \vee s)\right]du dr ds\nonumber\\
&=& \int_{t}^{T} \int_{s}^{T}\int_{r}^{T} 4 \alpha^2\ \sigma_0^2 e^{\alpha^2 u}1_{[0,u)}(r \vee s)du dr ds\nonumber\\
&=& 4 \alpha^2\ \sigma_0^2 \int_{t}^{T} \int_{s}^{T} \frac{e^{\alpha^2 T} - e^{\alpha^2 r}}{\alpha^2} dr ds\nonumber\\
&=& 4 \alpha^2\ \sigma_0^2 \int_{t}^{T} \frac{e^{\alpha^2 T}(T-s)}{\alpha^2} - \frac{e^{\alpha^2 T} - e^{\alpha^2 s}}{\alpha^4} ds\nonumber\\
&=& 4 \alpha^2\ \sigma_0^2 \left( \frac{e^{\alpha^2 T}(T-t)^2}{2\alpha^2} - \frac{e^{\alpha^2 T}(T-t)}{\alpha^4} + \frac{e^{\alpha^2 T} - e^{\alpha^2 t}}{\alpha^6} \right).\label{t3}
\end{eqnarray}
Here, we assume $t=0$ and apply the Taylor expansion,
\begin{eqnarray}
\eqref{t3}
&=& 
4\alpha^2 \sigma_0^2  \left(\frac{1+\alpha^2T}{2\alpha^2}T^2 - \frac{1+\alpha^2T + \frac{1}{2}\alpha^4T^2}{\alpha^4}T + \frac{\alpha^2T + \frac{1}{2}\alpha^4T^2 + \frac{1}{6}\alpha^6T^3}{\alpha^6}\right)+O(T^4)\nonumber\\
&=& 
\frac{2 \alpha^2\sigma_0^2}{3}T^3+O(T^4).
\end{eqnarray}
\begin{eqnarray}
T_4
&=& 
\int_{t}^{T}\int_{s}^{T}E_{t}\left[D_{s}^{W}\sigma _{r}^{2}\right] dr ds \nonumber\\
&=&
\int_{t}^{T}\int_{s}^{T}2\alpha E_{t}\left[ \sigma_0^2 e^{2\alpha W_r - \alpha^2 r}\right] dr ds \nonumber\\
&=&
\int_{t}^{T}\int_{s}^{T}2\alpha \sigma_0^2 e^{\alpha^2 r} dr ds \nonumber\\
&=&
2\alpha \sigma_0^2 \int_{t}^{T} \frac{e^{\alpha^2 T} - e^{\alpha^2 s}}{\alpha^2} ds \nonumber\\
&=&
2\alpha \sigma_0^2  \left(\frac{e^{\alpha^2 T}(T-t)}{\alpha^2} - \frac{e^{\alpha^2 T} - e^{\alpha^2 t}}{\alpha^4} \right).\label{t4}
\end{eqnarray}
Here, we assume $t=0$ and apply the Taylor expansion,
\begin{eqnarray}
\eqref{t4}
&=& 
2\alpha \sigma_0^2  \left(\frac{1+\alpha^2 T}{\alpha^2}T - \frac{\alpha^2T + \frac{1}{2}\alpha^4 T^2}{\alpha^4}\right) + O(T^3)\nonumber\\
&=& 
\alpha \sigma_0^2  \left(T^2 - \frac{1}{2}T^2\right) + O(T^3)\nonumber\\
&=& 
\frac{\alpha \sigma_0^2}{2}T^2+O(T^3).
\end{eqnarray}

Thus,
\begin{eqnarray}
\lefteqn{\lim_{T\rightarrow 0}\frac{I( 0,T,X_0,k_0^*) -E_{0}[v_{0}]} {T}}\nonumber \\
&\le& 
\lim_{T\rightarrow 0}\frac{3\rho^{2}}{8\sigma _{0}^{3}T^{4}} \frac{4 \alpha^2 \sigma_0^4}{3}T^4
- \lim_{T\rightarrow 0}\frac{\rho ^{2}}{{2\sigma _{0}^2T^{3}}}\frac{\alpha^2 \sigma_0^3}{3}T^3
- \lim_{T\rightarrow 0}\frac{\rho ^{2}}{2\sigma _{0}T^{3}} \frac{2 \alpha^2\sigma_0^2}{3}T^3
+ \lim_{T\rightarrow 0}\frac{\rho}{4T^{2}} \frac{\alpha \sigma_0^2}{2}T^2 + \lim_{T\to 0}o(T)\nonumber\\
&=& \frac{\rho \alpha \sigma_0^2}{8}.
\end{eqnarray}
Thus, ATMI is a lower bound of the volatility swap if $\rho<0$.
\end{proof}
\end{theorem}

\subsection{Zero-vanna implied volatility}
\begin{theorem} Consider the SABR model defined as in (\ref{themodel}) and (\ref{SABR}). Then, for all values of correlation $\rho$, the ZVIV is a lower bound for the volatility swap strike in the short time to maturity limit.
\begin{equation}
\lim_{T \to t} \, I(t,T,X_t,\hat{k}_t) \leq \lim_{T \to t} \,  E_{t}[v_{t}].
\end{equation}
\begin{proof}
From Theorem 4 in \cite{ARS},
\begin{eqnarray}
\label{lim1}
\lefteqn{\lim_{T\rightarrow t}\frac{I( t,T,X_t,\hat{k}_t) -E_{t}[v_{t}]} {(T-t)}}\nonumber \\
&=& \lim_{T\rightarrow t}\frac{3\rho^{2}}{8\sigma _{t}^{3}(T-t)^{4}}
E_{t}\left[\left(\int_{t}^{T}\int_{s}^{T}D_{s}^{W}\sigma _{r}^{2}dr ds\right)^{2} \right]\nonumber\\
&&-\lim_{T\rightarrow t}\frac{\rho ^{2}}{{2\sigma _{t}^2(T-t)^{3}}}
E_{t}\left[ \int_{t}^{T} \int_{s}^{T}D_{s}^{W}\sigma_{r} \int_{r}^{T}D_{r}^{W}\sigma _{u}^{2}du dr ds\right]\nonumber\\
&&-\lim_{T\rightarrow t}\frac{\rho ^{2}}{2\sigma _{t}(T-t)^{3}}
E_{t}\left[\int_{t}^{T} \int_{s}^{T}\int_{r}^{T}D_{s}^{W}D_{r}^{W}\sigma _{u}^{2}du dr ds\right]\nonumber\\
&=& \lim_{T\rightarrow t}\frac{3\rho^{2}}{8\sigma _{t}^{3}(T-t)^{4}}T_1 - \lim_{T\rightarrow t}\frac{\rho ^{2}}{{2\sigma _{t}^2(T-t)^{3}}} T_2 - \lim_{T\rightarrow t}\frac{\rho ^{2}}{2\sigma _{t}(T-t)^{3}} T_3.
\end{eqnarray}
Since $T_1$, $T_2$, $T_3$ are the same as those in the proof of Theorem \ref{correlatedATMI}, we obtain
\begin{eqnarray}
\lefteqn{\lim_{T\rightarrow 0}\frac{I( 0,T,X_0,\hat{k}_0) -E_{0}[v_{0}]} {T}}\nonumber \\
&\le& 
\lim_{T\rightarrow 0}\frac{3\rho^{2}}{8\sigma _{0}^{3}T^{4}} \frac{4 \alpha^2 \sigma_0^4}{3}T^4
- \lim_{T\rightarrow 0}\frac{\rho ^{2}}{{2\sigma _{0}^2T^{3}}}\frac{\alpha^2 \sigma_0^3}{3}T^3
- \lim_{T\rightarrow 0}\frac{\rho ^{2}}{2\sigma _{0}T^{3}} \frac{2 \alpha^2\sigma_0^2}{3}T^3 + \lim_{T\to 0}o(T)\nonumber\\
&=&0.
\end{eqnarray}
\end{proof}
\end{theorem}

\section{Numerical results}
In the previous sections it has been proved that in the short time to maturity limit the ZVIV bounds the volatility swap price from below. For longer time to maturities the numerical experiments in this section suggests that the ZVIV remains a lower bound. A proof of this conjecture, however, is currently not known. For the numerical experiments the volatility swap strike, the ZVIV and the ATMI are calculated for various values of volatility of volatility and time to maturity:
$$
\alpha \in \{0.5,1\}, \quad T \in \{0.5,1\}.
$$
The values of correlation are set in 0.1 increments from -1 to 1, and the initial value of the volatility is $\sigma_0 = 0.3$.
In order to calculate the volatility swap strike, the ZVIV and ATMI, $10^7$ simulations are carried out for each value of $\rho, \alpha$ and $T$. The results are depicted in Figures 1 to 4.

As can be deduced from the figures, and in line with the results of the previous sections and in \cite{ARS}, the ZVIV is indeed a more accurate approximation for the volatility swap price than the ATMI. Furthermore the impact of correlation on ZVIV is almost of second order. The term 'almost' is used since the ZVIV approximation is obtained by Taylor approximations in \cite{RA}. Thus there are still residual terms of $O(\rho)$ that remain in the ZVIV. The impact of these terms is apparent for longer time to maturities, significant nonzero values for correlation, and high volatility of volatility as in Figure 4. We can also clearly see in Figure 4 that for $\rho = 0$ the ZVIV is not exactly equal to the volatility swap strike. This is another manifestation of the fact that there are small residual terms of  $O(\rho)$ present in the ZVIV.

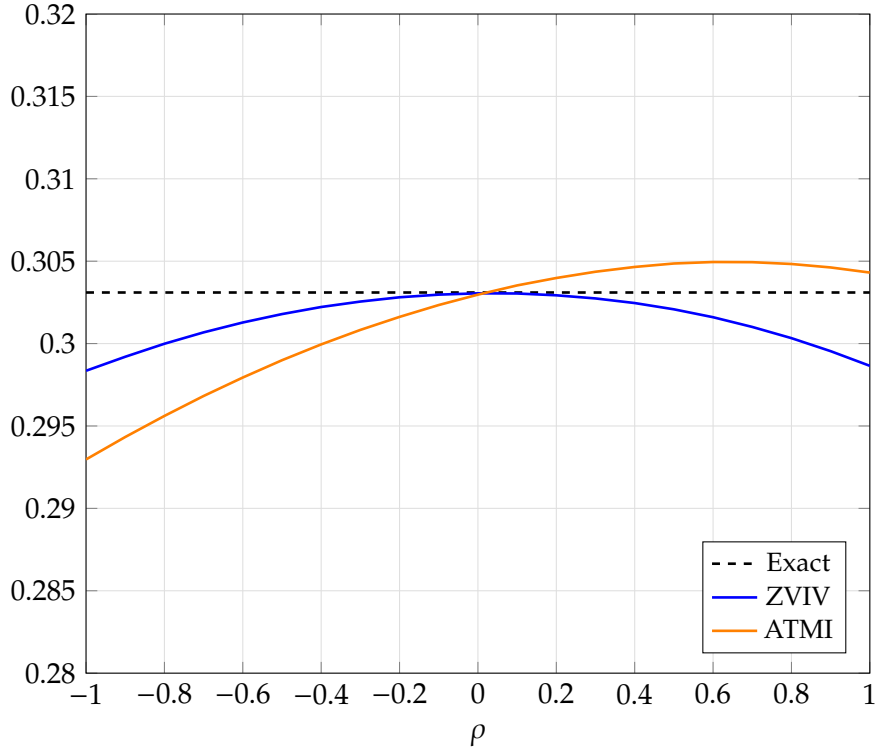
\begin{figure}
\centering
\begin{tikzpicture}
\begin{axis}[
        yticklabel style={/pgf/number format/fixed,
                     /pgf/number format/precision=3},
	legend pos= south east,
	legend style={font=\small},
	xlabel=$\rho$, 
	ymin = .28, ymax = .32,
	xmin = -1, xmax = 1,
	grid=both,
	minor grid style={gray!25},
	major grid style={gray!25},
	width=0.75 \linewidth,
	no marks]
	
\addplot[line width=1pt,dashed,color=black] 
	table[x=x, y=y1, col sep=tab]{fig1.txt};
\addlegendentry{Exact};	
	
\addplot[line width=1pt,solid,color=blue] 
	table[x=x, y=y2, col sep=tab]{fig1.txt};
\addlegendentry{ZVIV};	

\addplot[line width=1pt,solid,color=orange] 
	table[x=x, y=y3, col sep=tab]{fig1.txt};
\addlegendentry{ATMI};

\end{axis}
\end{tikzpicture}
\caption{$\sigma_0 = 0.3, \; \alpha = 0.5, \; T = 0.5$}
\end{figure}

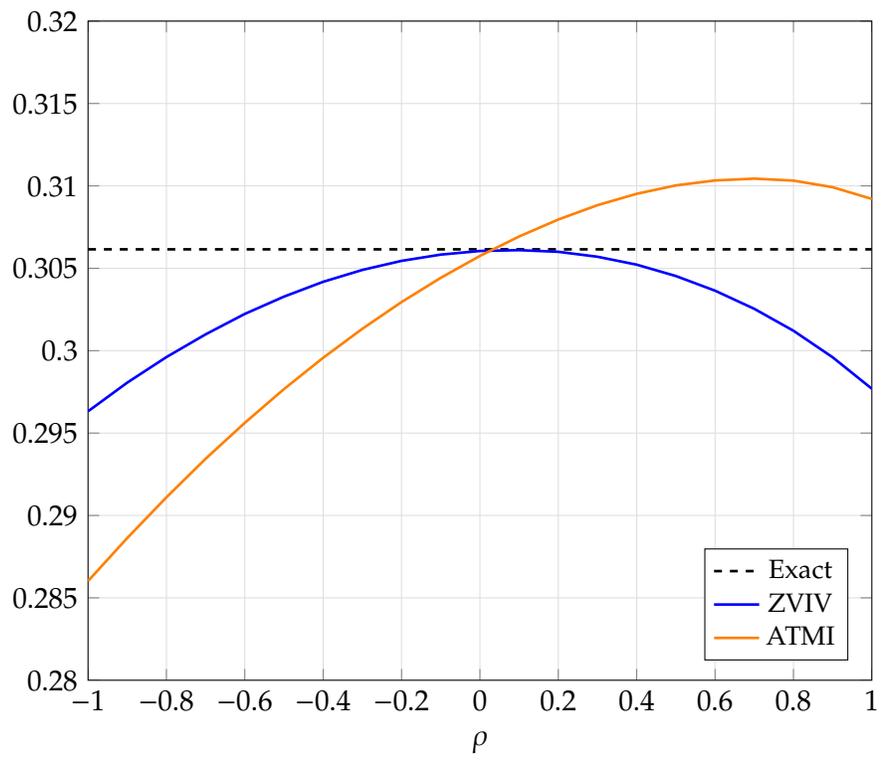
\begin{figure}
\centering
\begin{tikzpicture}
\begin{axis}[
        yticklabel style={/pgf/number format/fixed,
                     /pgf/number format/precision=3},
	legend pos= south east,
	legend style={font=\small},
	xlabel=$\rho$, 
	ymin = .28, ymax = .32,
	xmin = -1, xmax = 1,
	grid=both,
	minor grid style={gray!25},
	major grid style={gray!25},
	width=0.75 \linewidth,
	no marks]
	
\addplot[line width=1pt,dashed,color=black] 
	table[x=x, y=y1, col sep=tab]{fig2.txt};
\addlegendentry{Exact};	
	
\addplot[line width=1pt,solid,color=blue] 
	table[x=x, y=y2, col sep=tab]{fig2.txt};
\addlegendentry{ZVIV};	

\addplot[line width=1pt,solid,color=orange] 
	table[x=x, y=y3, col sep=tab]{fig2.txt};
\addlegendentry{ATMI};

\end{axis}
\end{tikzpicture}
\caption{$\sigma_0 = 0.3, \; \alpha = 0.5, \; T = 1$}
\end{figure}

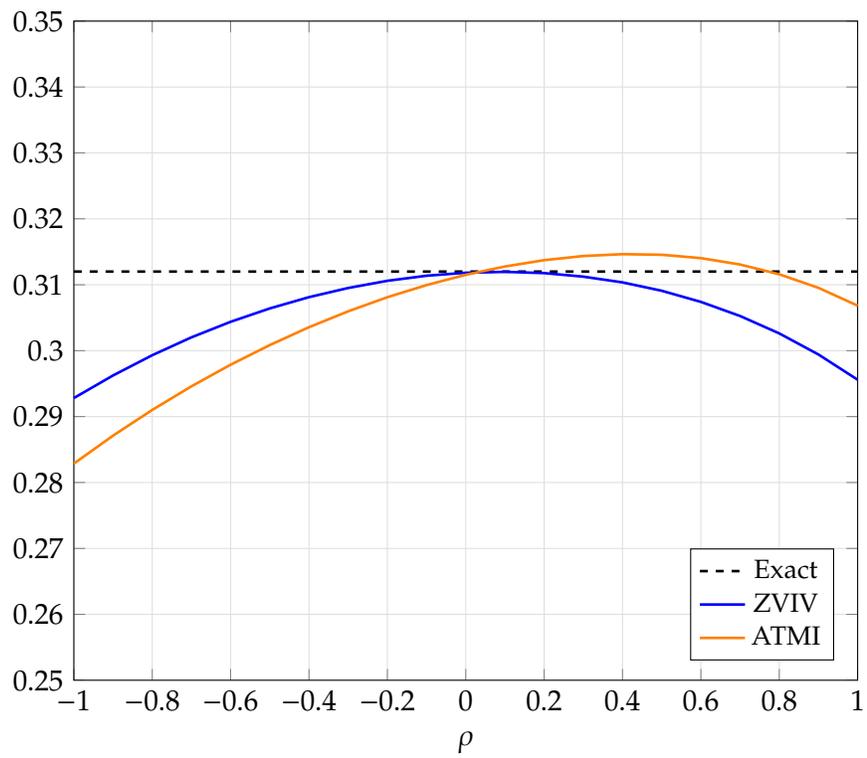
\begin{figure}
\centering
\begin{tikzpicture}
\begin{axis}[
        yticklabel style={/pgf/number format/fixed,
                     /pgf/number format/precision=3},
	legend pos= south east,
	legend style={font=\small},
	xlabel=$\rho$, 
	ymin = .25, ymax = .35,
	xmin = -1, xmax = 1,
	grid=both,
	minor grid style={gray!25},
	major grid style={gray!25},
	width=0.75 \linewidth,
	no marks]
	
\addplot[line width=1pt,dashed,color=black] 
	table[x=x, y=y1, col sep=tab]{fig3.txt};
\addlegendentry{Exact};	
	
\addplot[line width=1pt,solid,color=blue] 
	table[x=x, y=y2, col sep=tab]{fig3.txt};
\addlegendentry{ZVIV};	

\addplot[line width=1pt,solid,color=orange] 
	table[x=x, y=y3, col sep=tab]{fig3.txt};
\addlegendentry{ATMI};

\end{axis}
\end{tikzpicture}
\caption{$\sigma_0 = 0.3, \; \alpha = 1, \; T = 0.5$}
\end{figure}

\begin{figure}
\centering
\begin{tikzpicture}
\begin{axis}[
        yticklabel style={/pgf/number format/fixed,
                     /pgf/number format/precision=3},
	legend pos= south east,
	legend style={font=\small},
	xlabel=$\rho$, 
	ymin = .25, ymax = .35,
	xmin = -1, xmax = 1,
	grid=both,
	minor grid style={gray!25},
	major grid style={gray!25},
	width=0.75 \linewidth,
	no marks]
	
\addplot[line width=1pt,dashed,color=black] 
	table[x=x, y=y1, col sep=tab]{fig4.txt};
\addlegendentry{Exact};	
	
\addplot[line width=1pt,solid,color=blue] 
	table[x=x, y=y2, col sep=tab]{fig4.txt};
\addlegendentry{ZVIV};	

\addplot[line width=1pt,solid,color=orange] 
	table[x=x, y=y3, col sep=tab]{fig4.txt};
\addlegendentry{ATMI};

\end{axis}
\end{tikzpicture}
\caption{$\sigma_0 = 0.3, \; \alpha = 1, \; T = 1$}
\end{figure}
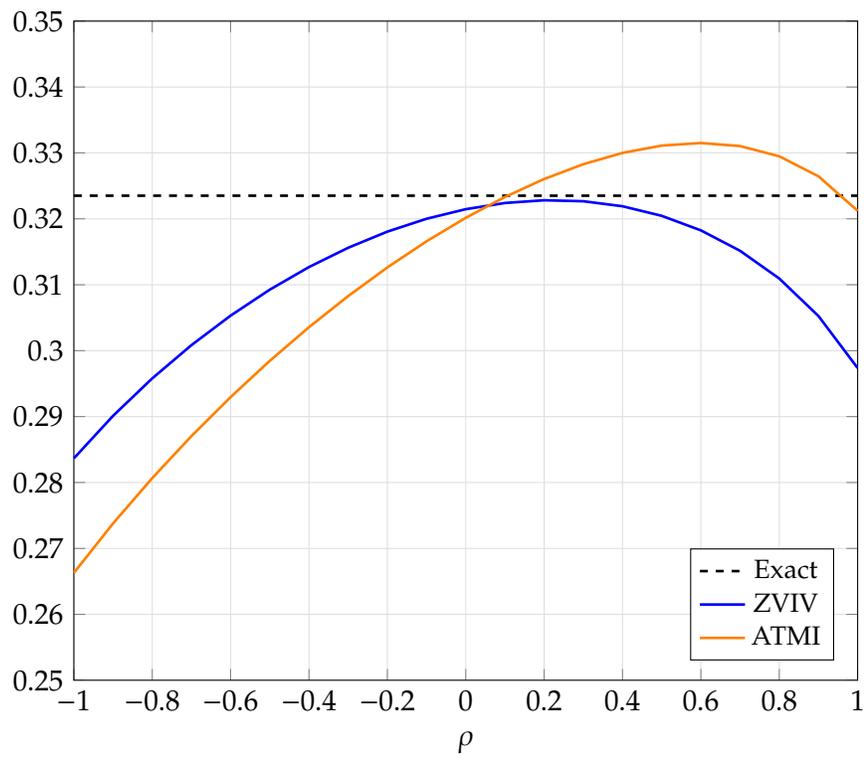

\section{Conclusion}
It has been shown that in the conditionally lognormal SABR model, the ZVIV is a lower bound for the volatility swap strike for all values of correlation in the short time to maturity limit. This property, in addition to the results in \cite{ARS}, further cements the ZVIV as a more accurate model-free approximation for the volatility swap strike than the ATMI.

The results in this paper is a first step in analyzing the question for which models the ZVIV is a lower bound for the volatility swap strike. Further research into the ZVIV as a lower bound for the volatility swap strike will have to includes analyses along the dimensions of the Hurst parameter, correlation parameter and time to maturity.


\begin{thebibliography}{9}





\bibitem{AS} Al\`{o}s, E., and Shiraya, K. ``Estimating the Hurst parameter from short term volatility swaps: a Malliavin calculus approach.'' Finance and Stochastics 23.2 (2019): 423-447. doi: 10.1007/s00780-019-00384-5

\bibitem{ARS} Al\`{o}s, E., Rolloos, F., and Shiraya, K. ``On the difference between the volatility swap strike and the zero vanna implied volatility'' SIAM Journal on Financial Mathematics 12 (2) (2021), doi: 10.1137/20M134722X



\bibitem{CL1} Carr, P. and Lee, R. ``Robust replication of volatility derivatives,'' PRMIA award for Best Paper in Derivatives, MFA 2008 Annual Meeting, (2008), doi:10.2139/ssrn.1108429. 


\bibitem{CR} Comte, F. and Renault, E.: ``Long memory in continuous-time stochastic volatility models,''  Math. Financ. 8, (1998), 291-323, doi: 10.1111/1467-9965.00057







\bibitem{FG} Friz, P., and Gatheral, J. ``Valuation of volatility derivatives as an inverse problem,'' Quantitative Finance 5 (6), (2005), doi: 531-542. 10.1080/14697680500362452


\bibitem{GJR} Gatheral, J., Jaisson, T. and Rosenbaum, M.  Volatility is rough, Quantitative Finance, 18:6 (1998), 933-949. doi: 10.1080/14697688.2017.1393551



\bibitem{RA} Rolloos, F., and Arslan, M. ``Taylor-made volatility swaps,'' Wilmott, January (2017), 56-61, doi: 10.1002/wilm.10566



\end{thebibliography}
\end{document}